\documentstyle{article}

\def\be{\begin{eqnarray}}
\def\ee{\end{eqnarray}}
\def\nn{\nonumber}

\hoffset= - 1.0in         % switch off for draft style
\def\pA{{\tilde{\cal A}}}
\def\npA{{\cal A}}
\def\pF{{\tilde{\cal F}}}

\def\pD{{\tilde{\cal D}}}

\def\bj{\,\bar{\! j}}

\begin{document}

\hfill ITEP/TH-16/08

\bigskip

\centerline{\Large{On the Problem of Multiple M2 Branes}}

\bigskip

\centerline{A.Morozov}

\bigskip

\centerline{\it ITEP, Moscow, Russia}

\bigskip

\centerline{ABSTRACT}

\bigskip

{\footnotesize
A simplified version of $3d$ BL theory is considered,
which allows any number $N$ of $M2$ branes in $d=11$.
The underlying $3$-algebra structure is provided
by degenerate $U(N)$ Nambu bracket
$[X,Y,Z] = {\rm tr}(X)\cdot [Y,Z] + {\rm tr}(Y)\cdot [Z,X] +
{\rm tr}(Z)\cdot [X,Y]$, the corresponding
$f^{abcd}$ is not totally antisymmetric and the ${\cal N}=8$
supersymmetry of the action remains to be checked.
All the fields, including auxiliary non-propagating
gauge fields, are in adjoint representation of $SU(N)$
and the only remnant of $3$-algebra structure is an
octuple of gauge singlets, acquiring vacuum expectation
value in transition to $D2$ branes in $d=10$.
}

\bigskip

\bigskip

\tableofcontents

\section{Introduction}

In \cite{JS} John Schwarz explicitly formulated a problem
to find the world-volume $3d$ non-Abelian action
with $OSp(8|4)$ symmetry,
including the ${\cal N}=8$ supersymmetry and
conformal invariance,
which could be used in description of $N$ parallel fundamental
$M2$ branes embedded into $11d$ space-time.
It was explained in \cite{JS} that $3d$ gauge fields in this
theory should be non-dynamical and governed by a Chern-Simons
action.
After important work \cite{BH},
the problem was finally  resolved by Jonathan Bagger and
Neil Lambert (BL) in a series of papers \cite{BL1,BL2,BL3}
and was further elaborated on in \cite{G1,G2,MP,BLS,BTT,MVR}.
The BL construction is based on non-trivial
$3$-algebra structure and involves a new kind of gauge
fields, which do not belong to adjoint representation of
naive gauge group $U(N)$.
Moreover, additional antisymmetry requirement imposed in
\cite{BL2} on the $3$-algebra structure constants $f^{abcd}$,
leaves only one non-trivial example $f^{abcd} \sim \epsilon^{abcd}$,
restricting the choice of the "gauge group" to $U(2)$
(or, perhaps, $SO(4)$).
A possible way out was actually suggested in
\cite{G2} and \cite{MP},
where $f^{abcd}$ were linked to the structure constants
$f^{abc}$ of the ordinary gauge group $G$, say, $G=U(N)$:
$f^{abc0} = f^{abc}$.
As already noted at the very end of \cite{G2}, if one
relaxes the unnecessary antisymmetry constraint,
this identification essentially implies the use of the
standard quantum Nambu $3$-bracket for ordinary matrices
\cite{ALMY}
\be
[X,Y,Z] = {\rm tr}(X)\cdot [Y,Z] + {\rm tr}(Y)\cdot [Z,X] +
{\rm tr}(Z)\cdot [X,Y]
\label{Nabra}
\ee
in the role of BL $3$-algebra structure.

In what follows we explicitly describe this simplified
version of BL construction for arbitrary gauge group $G$,
including $G=U(N)$ relevant for the stack of $N$ $M2$ branes.
The non-trivial BL gauge fields reduce to the pair of
ordinary adjoint auxiliary fields, one gauge and one not,
very much in the spirit of \cite{MP}.
We begin in s.\ref{toy} and s.\ref{fref} from reminding
respectively the
${\cal N}=1$ and ${\cal N}=8$ SUSY non-gauged $3d$ actions
from \cite{JS} with adjoint octuplet
$(\phi^I_{i\bj}, \psi^A_{i\bj})$,
$\ I=1 \ldots 8$, $A=1\ldots 8$, $\ i,j=1\ldots N$.
Then in s.\ref{nalin} its simplest gauged generalization
is considered, with two auxiliary vector fields:
gauge $A^\mu_{i\bj}$ and additional adjoint $B^\mu_{i\bj}$.
A non-linear potential {\it a la} BL is introduced in
s.\ref{prepo}, involving additional octuplet of gauge singlets
$(\varphi^I,\chi^A)$.
SUSY invariance of this action is addressed in s.\ref{var},
where some representative but non-exhaustive examples are
presented.
In s.\ref{BLa} we briefly remind the main points of original
BL construction for $M2$ in $d=11$ and the way \cite{MP}
it reduces to $D2$ in $d=10$.
Finally in s.\ref{BLasi} we demonstrate that the substitution
of  (\ref{Nabra}) converts this BL action into the simple
formula (\ref{Act}) from our s.\ref{prepo}, which is the
main suggestion of the present paper.

\section{Non-gauged ${\cal N}=1$ SUSY action in $3d$ \label{toy}}

Consider a {\it real} (Grassmann-valued) $3d$ spinor
$\psi = \left(\begin{array}{c}\psi_+ \\ \psi_- \end{array}\right)$.
It exists if space-time signature is $(-++)$
and three {\it real}-valued
gamma-matrices are $\gamma_0 = i\sigma_2 =
\left(\begin{array}{cc} 0&1\\-1&0 \end{array}\right)$,
$\gamma_1=
\sigma_1 = \left(\begin{array}{cc} 0&1\\1&0 \end{array}\right)$
and $\gamma_2 =
\sigma_3=\left(\begin{array}{cc} 1&0\\0&-1 \end{array}\right)$.
Then the Dirac Lagrangian
\be
{\cal L}_{Dir} = i\bar\psi \hat\partial \psi
= -\psi^\dagger \sigma_2
\Big(-i\sigma_2 \partial_0 + \sigma_1\partial_1 +
\sigma_3\partial_2\Big)\psi =
i\psi^\dagger\Big(\partial_0 + \sigma_3\partial_1
- \sigma_1\partial_2\Big) = \nn \\ =
(\psi_+,\psi_-)\left(\begin{array}{cc}
i(\partial_0 +\partial_1)& -i\partial_2 \\
-i\partial_2 & i(\partial_0 - \partial_1)
\end{array}\right)
\left(\begin{array}{c}\psi_+ \\ \psi_- \end{array}\right)
= \nn \\ =
i\psi_+(\partial_0 +\partial_1)\psi_+
+i\psi_-(\partial_0 -\partial_1)\psi_-
- i\psi_+\partial_2\psi_- - i\psi_-\partial_2\psi_+
\ee
is non-vanishing, because of the anticommuting nature of $\psi$-fields.
It is related to bosonic Lagrangian
\be
{\cal L}_{bos} = -\partial^\mu\phi\partial_\mu\phi
= (\partial_0\phi)^2 - (\partial_1\phi)^2 - (\partial_2\phi)^2
\ee
by an elementary supersymmetry transformation
\be
\delta \phi = i\bar\psi\varepsilon =
(\psi_+,\psi_-) i\sigma_2
\left(\begin{array}{c}\varepsilon_+ \\ \varepsilon_-
\end{array}\right) = \psi_+\varepsilon_- - \psi_-\varepsilon_+
= %\nn \\ =
\varepsilon_+\psi_- - \varepsilon_-\psi_+ =
i\bar\varepsilon \psi; \nn \\
\delta \psi = \left(\begin{array}{c}\delta\psi_+ \\
\delta\psi_- \end{array}\right)
= -\hat\partial\phi \varepsilon =
\Big(i\sigma_2\partial_0\phi
- \sigma_1\partial_1\phi - \sigma_3\partial_2\phi\Big)
\left(\begin{array}{c}\varepsilon_+ \\ \varepsilon_-
\end{array}\right)
\ee
with constant infinitesimal  spinor $\varepsilon$, so that
\be
\delta{\cal L}_{bos} + \delta{\cal L}_{Dir} =
\partial_\mu \Big(-i\bar\psi \partial_\mu\phi \varepsilon\Big)
\ee
and the action
\be
\int\Big({\cal L}_{bos} + {\cal L}_{Dir}\Big)d^3x
\ee
remains invariant.

\section{Non-gauged ${\cal N}=8$ SUSY action in $3d$ \label{fref}}

The number of fields can be easily increased: just take
$M$ copies of $\phi$ and $\psi$.
What is non-trivial, for the special value of $M=8$
the number of supersymmetries can also be increased to $M$.
This can be done by using the special {\it triality}
relation between three different $8$-dimensional representations
of $SO(8)$ (or, what is essentially the same, triality relation
in octonionic algebra).
For reasons explained in \cite{BLS} we separate $3d$ and $8d$
gamma-matrices
and also denote the scalar fields by $\phi^I$ rather than $X^I$.

{\bf Fields:} Scalars $\phi^I$ and
$3d$ Majorana spinors $\psi_A$,
with $I$ and $A$ labeling components of $V_8$ and $S^+_8$
-- the vector and spinor
representations of $SO(8)$.
$4d$ spinor indices are suppressed.

{\bf ${\cal N}=8$ SUSY transformation}:
\be
\delta \phi^I = i\bar\varepsilon^{\dot A}
\Gamma^I_{\dot A A} \psi^A =
i\bar\psi^A \Gamma^I_{A\dot A} \varepsilon^{\dot A}, \nn \\
\delta \psi_A = -\hat\partial \phi^I
\Gamma^I_{A\dot A}\varepsilon^{\dot A}
\label{phipsitra0}
\ee
Here  $\varepsilon = \{\varepsilon^{\dot A}\}$ belongs
to the second spinor representation $S^-_8$ of $SO(8)$,
the $8\times 8\times 8$  real-valued structure constants
$\Gamma^I_{A\dot A}$
define the {\it triality} relation between the three $8$-dimensional
representations of $SO(8)$.
They are $8\times 8$ off-diagonal blocks in $16\times 16$
gamma-matrices in $8d$:
$$
\gamma^I = \left(\begin{array}{cc} 0 & \Gamma^I_{A\dot A} \\
\Gamma^I_{\dot A A} & 0 \end{array}\right)
= \left(\begin{array}{cc} 0 & \Gamma^I \\
\check\Gamma^I & 0 \end{array}\right)
\ \ \ \ {\rm and}\ \ \ \
\gamma^I\gamma^J+\gamma^J\gamma^I = 2\delta^{IJ}
\Longrightarrow  \Gamma^I\check \Gamma^J + \Gamma^J\check\Gamma^I =
2\delta^{IJ}
$$
where $\check\Gamma$ denotes transposed matrix.
In particular basis $\Gamma^8_{\alpha\beta} = \delta_{\alpha\beta}$
while for $i,j,k=1\ldots 7$ we have
$\Gamma^i_{j8} = -\Gamma^i_{8j} = \delta_{ij}$
and $\Gamma^i_{jk} = c_{ijk}$ where $c_{ijk} = -c_{ikj} = c_{jki}$
are octonionic structure constants, non-vanishing for the following
triples:
\be
c_{124} = c_{137} = c_{156} = c_{235}
= c_{267} = c_{346} = c_{457} = 1
\ee
Explicitly in this basis

{\footnotesize \centerline{ $
\Gamma^1 = \left(\begin{array}{cccccccc}
0&0&0&0&0&0&0&1\\
0&0&0&1&0&0&0&0\\
0&0&0&0&0&0&1&0\\
0&-1&0&0&0&0&0&0\\
0&0&0&0&0&1&0&0\\
0&0&0&0&-1&0&0&0\\
0&0&-1&0&0&0&0&0\\
-1&0&0&0&0&0&0&0
\end{array}\right) \ \ \
\Gamma^2 = \left(\begin{array}{cccccccc}
0&0&0&-1&0&0&0&0\\
0&0&0&0&0&0&0&1\\
0&0&0&0&1&0&0&0\\
1&0&0&0&0&0&0&0\\
0&0&-1&0&0&0&0&0\\
0&0&0&0&0&0&1&0\\
0&0&0&0&0&-1&0&0\\
0&-1&0&0&0&0&0&0
\end{array}\right) \ \ \
\Gamma^3 = \left(\begin{array}{cccccccc}
0&0&0&0&0&0&-1&0\\
0&0&0&0&-1&0&0&0\\
0&0&0&0&0&0&0&1\\
0&0&0&0&0&1&0&0\\
0&1&0&0&0&0&0&0\\
0&0&0&-1&0&0&0&0\\
1&0&0&0&0&0&0&0\\
0&0&-1&0&0&0&0&0
\end{array}\right)
$}

\centerline{$
\Gamma^4 = \left(\begin{array}{cccccccc}
0&1&0&0&0&0&0&0\\
-1&0&0&0&0&0&0&0\\
0&0&0&0&0&-1&0&0\\
0&0&0&0&0&0&0&1\\
0&0&0&0&0&0&1&0\\
0&0&1&0&0&0&0&0\\
0&0&0&0&-1&0&0&0\\
0&0&0&-1&0&0&0&0
\end{array}\right) \ \ \
\Gamma^5 = \left(\begin{array}{cccccccc}
0&0&0&0&0&-1&0&0\\
0&0&1&0&0&0&0&0\\
0&-1&0&0&0&0&0&0\\
0&0&0&0&0&0&-1&0\\
0&0&0&0&0&0&0&1\\
1&0&0&0&0&0&0&0\\
0&0&0&1&0&0&0&0\\
0&0&0&0&-1&0&0&0
\end{array}\right)\ \ \
\Gamma^6 = \left(\begin{array}{cccccccc}
0&0&0&0&1&0&0&0\\
0&0&0&0&0&0&-1&0\\
0&0&0&1&0&0&0&0\\
0&0&-1&0&0&0&0&0\\
-1&0&0&0&0&0&0&0\\
0&0&0&0&0&0&0&1\\
0&1&0&0&0&0&0&0\\
0&0&0&0&0&-1&0&0
\end{array}\right) $} $$
\Gamma^7 = \left(\begin{array}{cccccccc}
0&0&1&0&0&0&0&0\\
0&0&0&0&0&1&0&0\\
-1&0&0&0&0&0&0&0\\
0&0&0&0&1&0&0&0\\
0&0&0&-1&0&0&0&0\\
0&-1&0&0&0&0&0&0\\
0&0&0&0&0&0&0&1\\
0&0&0&0&0&0&-1&0
\end{array}\right) \ \ \
\Gamma^8 = \left(\begin{array}{cccccccc}
1&0&0&0&0&0&0&0\\
0&1&0&0&0&0&0&0\\
0&0&1&0&0&0&0&0\\
0&0&0&1&0&0&0&0\\
0&0&0&0&1&0&0&0\\
0&0&0&0&0&1&0&0\\
0&0&0&0&0&0&1&0\\
0&0&0&0&0&0&0&1
\end{array}\right)
$$ }
Indices $I$, $A$ and $\dot A$ are raised and lowered with the help
of invariant  metrics $g_{IJ}$, $g_{AB}$ and $g_{\dot A\dot B}$.

{\bf Invariant action}:
Lagrangian
\be
{\cal L}_{{\rm free}} =
- \partial^\mu\phi^I\partial_\mu\phi^I +
i\bar\psi^A\hat\partial \psi^A
\label{Lfree}
\ee
changes under (\ref{phipsitra0}) by total derivative:
\be
-2\partial^\mu \phi^I \partial_\mu\Big(
i\bar\psi^A \Gamma^I_{A\dot A} \varepsilon^{\dot A}\Big)
\longrightarrow +2i\partial^2\phi^I
\Big(\bar\psi^A \Gamma^I_{A\dot A} \varepsilon^{\dot A}\Big), \nn \\
2i\bar\psi^A \hat\partial \Big(-\hat\partial \phi^I
\Gamma^I_{A\dot A}\epsilon^{\dot A}\Big)
= -2i\partial^2\phi^I
\Big(\bar\psi^A \Gamma^I_{A\dot A} \varepsilon^{\dot A}\Big)
\ee
so that the action
\be
\int {\cal L}_{{\rm free}} d^3x
\ee
remains invariant.

\section{Gauged ${\cal N}=8$ SUSY action in $3d$ \label{nalin}}

If $\phi$ and $\psi$ are promoted to $N\times N$ matrices,
or to elements of adjoint representation of any other group $G$,
$\phi^I_a$, $\psi^A_a$, $a=1\ldots {\rm dim}(G)$, the action
acquires a global $G$-symmetry, which can further be gauged
by introduction of the gauge field $A^\mu_a$:
\be
{\cal L}_{\rm {kin}} =
-(D^\mu\phi^I)_a(D_\mu\phi^I)_a + i\bar\psi^A_a(\hat D\psi^A)_a
\ee
with
\be
D_\mu^{ab} = \eta^{ab}\partial_\mu + f^{abc}A_\mu^c
\ee
where $f^{abc}$ are the structure constants of $G$,
satisfying Jacobi identity. Indices $a$ are raised and
lowered with the help of the Killing metric $\eta_{ab}$.
However, the action
\be
\int {\cal L}_{\rm{kin}}d^3x
\ee
is not invariant under the ${\cal N}=8$ SUSY transformations
\be
\delta \phi^I_a = i\bar\varepsilon^{\dot A}
\Gamma^I_{\dot A A} \psi^A_a =
i\bar\psi^A_a \Gamma^I_{A\dot A} \varepsilon^{\dot A}, \nn \\
\delta \psi_A^a = -(\hat D \phi_I)^a
\Gamma^I_{A\dot A}\varepsilon^{\dot A}
\label{phipsitrana0}
\ee
because $(\hat D^2 - D^2)_{ab} = \frac{1}{2}
f_{abc}F_c^{\mu\nu}\sigma_{\mu\nu}
\neq 0$,
where $F_{\mu\nu}^a = \partial_\mu A_\nu^a - \partial_\nu A_\mu^a
+ f^{abc}A_\mu^bA_\nu^c$ and $\sigma_{\mu\nu} = \frac{1}{2}
[\gamma_\mu,\gamma_\nu] = \epsilon_{\lambda\mu\nu}\gamma^\lambda$.
The variation of the action is
\be
\delta \int {\cal L}_{\rm{kin}}d^3x = i\epsilon^{\lambda\mu\nu}
\left(\int \bar\psi^A_a F_{\mu\nu}^c\gamma^\lambda
\phi^I_bd^3x\right)
f^{abc}\Gamma^I_{A\dot A}\varepsilon^{\dot A}
\ee
and can be easily compensated by adding a Chern-Simons-like
term
\be
S_{CS} = \epsilon^{\lambda\mu\nu}\int B^a_\lambda F^a_{\mu\nu}d^3x
\ee
with additional auxiliary pseudovector field $B^a_\mu$ in adjoint
representation of $G$, which varies under ${\cal N}=8$ SUSY
transformation:
\be
\delta B^c_\mu =
-i\Big(\bar\psi^A_a \gamma^\lambda \varepsilon^{\dot A}\Big)
\phi^I_b f^{abc}\Gamma^I_{A\dot A},
\label{Btrana0}
\ee
while
\be
\delta A^c_\mu = 0
\label{Atrana0}
\ee
Note that there is no $B^3$ term in Chern-Simons action.
Therefore one can say that this auxiliary field $B$
works as Lagrange multiplier,
nullifying the gauge curvature $F_{\mu\nu}^c$ on-shell,
what makes it the theory essentially linear in
flat-connection background.
Still this theory
\be
\int {\cal L}_{{\rm kin}} d^3x + S_{CS},
\label{act}
\ee
though rather trivial, possesses all the desired properties:
${\cal N}=8$ supersymmetry, conformal invariance (at least
classical) and it is also $P$-invariant, provided $B_\mu^c$
is a pseudovector.

\section{Non linear $U(N)$ gauged ${\cal N}=8$ SUSY action in $3d$
\label{prepo}}

The theory can be made more interesting by introducing
non-linear potential and accompanying $\psi^2$ terms.
This can be done for any group $G$, but we write it in
terms of the most interesting $G=U(N)$, to avoid repetition
of formulas from the previous section.
Adjoint representation of $SU(N)$ can be described as
anti-Hermitian traceless $N\times N$ matrices, so that the
single index $a=1,\ldots,N^2-1$ turns into a pair of
indices $a=(i\bj)$, $i,j=1\ldots N$, the structure constants
$f^{abc}$ are induced by matrix commutators and Killing metric
-- by matrix trace.
Still, formulation in terms of $f^{abc}$ in the previous section
is also useful not only for more complicated Lie algebras, it also
allows to neglect details, associated with the complex-valuedness
of anti-Hermitian matrices.
One can also reduce $SU(N)$ to $SO(N)$, represented by
real antisymmetric $N\times N$ matrices, to fully avoid this
kind of problems.

Non-linearization turns to be related to the new $U(1)$
octuplet $(\varphi^I,\chi_A)$, it can be
associated with the unit matrix in $U(N)$, but in this section
we treat these fields simply as gauge singlets, without indices
$i\bj$ at all.

{\bf Fields:}

\vspace{+0.2cm}
\begin{tabular}{cccccc}
scalars & $\phi^I_{i\bj}$,& $\varphi^I$
& with & $I=1,\dots,8$, & $i,\bj=1,\ldots,N$ \\
&&&&&\\
spinors &$\psi^A_{i\bj}$,& $\chi_A$
&with &$A=1,\dots,8$, &$i,\bj=1,\ldots,N$\\
&&&&&\\
vectors &$A^\mu_{i\bj}$, &$B^\mu_{i\bj}$
&with& $\mu = 0,1,2,3$,& $i,\bj=1,\ldots,N$
\end{tabular}

\bigskip

{\bf Lagrangian:}
$$
- {\rm tr} ({\cal D}_\mu\phi^I)^2 +
i\,{\rm tr}\, \bar\psi^A \hat {\cal D} \psi^A
+\epsilon^{\lambda\mu\nu} {\rm tr}\ F_{\mu\nu}B_\lambda -
$$\vspace{-0.6cm}
\be
- (\partial_\mu\varphi^I)^2 + i\bar\chi^I\hat\partial\chi^I
+ 2i\varphi^I {\rm tr}\Big(\phi^J[\bar\psi^A,\psi^B]\Big)
\Gamma^{IJ}_{AB}
+ i\,{\rm tr} \Big( [\phi^I,\phi^J]\bar\psi^A\Big)\chi^B
\Gamma^{IJ}_{AB}
+ \sum_{K\neq I,J}
(\varphi^K)^2 {\rm tr}\left([\phi^I,\phi^J]\right)^2
\label{Act}
\ee
Sums are taken over repeated indices of all kinds.
The long derivatives here are
\be
({\cal D}^\mu\phi)_{i\bj}^I = \partial^\mu\phi^I_{i\bj}
+ [A^\mu,\phi^I]_{i\bj} + B^\mu_{i\bj}\varphi^I, \nn \\
({\cal D}^\mu\psi)_{i\bj}^A = \partial^\mu\psi^A_{i\bj}
+ [A^\mu,\psi^A]_{i\bj} + 2B^\mu_{i\bj}\chi^A
\label{lode}
\ee

\bigskip

{\bf $N=8$ SUSY transformations:}

\be
\delta\varphi^I = i\bar\varepsilon^{\dot A}
\Gamma^I_{A\dot A}\chi^A,\nn\\
%= (\bar\varepsilon\Gamma^I\chi)
\delta\chi_A =
-\hat\partial\varphi^I\Gamma^I_{A\dot A}\varepsilon^{\dot A}, \nn \\
%= (\hat\partial\varphi\Gamma\varepsilon)_A
\nn \\
\delta\phi^I_{i\bj} =
i\varepsilon^{\dot A} \Gamma^I_{A\dot A}\psi^A_{i\bj},\nn\\
(\delta\psi_A)_{i\bj} = -
(\hat{\cal D}\phi^I)_{i\bj}\Gamma^I_{A\dot A}\varepsilon^{\dot A}
- [\phi^I,\phi^J]_{i\bj}\varphi^K
\Gamma^{IJK}_{A\dot A}\varepsilon^{\dot A}, \nn \\
\nn \\
\delta A_{i\bj}^\lambda = \left(\varphi^I \bar\psi^A_{i\bj}
+ \phi^I_{i\bj} \bar\chi^A\right)
\gamma^\lambda\Gamma^I_{A\dot A}\varepsilon^{\dot A}, \nn \\
\delta B^\lambda_{i\bj} =
[\bar\psi^A,\phi^I]_{i\bj} \gamma^\lambda
\Gamma^I_{A\dot A}\varepsilon^{\dot A}
\ee

\bigskip

The "linear" action (\ref{act}) -- the first line in (\ref{Act})
--  and the SUSY
transformations
(\ref{phipsitrana0}), (\ref{Btrana0}), (\ref{Atrana0})
from the previous section
are reproduced if we put all $\varphi^I=\chi^A=0$,
what is a consistent {\it reduction} of the theory.

Instead one can ascribe an average value to the gauge singlet
$\varphi$, say
\be
\Big<\varphi^{I=8}\Big> = g_{YM}
\label{vev}
\ee
then the $B$ field acquires a "mass term" from
${\rm tr}\Big((D\phi)^8\Big)^2 \rightarrow
g_{YM}^2 {\rm tr}(B_\mu)^2$ and integrating $B$ out
one obtains a kinetic term
$\frac{1}{g^2_{YM}}{\rm tr} F_\mu\nu^2$
for the gauge field.
This is the $M2 \rightarrow D2$ transition, described in
\cite{MP} for the general BL theory.
Of course, eq.(\ref{vev}) is inconsistent with
equations of motion and is not an allowed VEV in
the theory (\ref{Act}).
One can cure this by adding non-trivial
potential to the field $\varphi$, what necessarily breaks
a part of supersymmetry as well as conformal invariance
 -- as required in transition to $D2$ branes.

\section{On SUSY invariance of (\ref{Act})
\label{var}}

Detailed check of invariance of (\ref{Act}) is somewhat
sophisticated and remains beyond the scope of the present
paper, we give in this section only some illustrative examples.
The real problem is of course the lack of superfield or
any other explicitly supersymmetric formulation.
Non-linear terms in BL action are not expressed even through
an ${\cal N}=1$ superpotential, i.e. do not look like
$\ \left(\frac{\partial W}{\partial\phi^I}\right)^2 +
\bar\psi^I\!\!\frac{\partial^2 W}{\partial\phi^I\partial\phi^J}
\psi^J\ $ for some $W(\phi)$, though they look surprisingly close
to this for a theory with highly extended supersymmetry
-- what can serve as additional inspiration for the search of
invariant formulations.
We remind the "natural" logic of invariance check for BL
action in the next s.\ref{BLa}, while here we directly collect
the terms of {\it some}
different structures in the variation of (\ref{Act}).
Of primary importance is variation of the potential,
since it has the open dependence on the matric $h_{ab}$
and thus can cause problems for non-antisymmetric $f^{abcd}$.

\subsection{$\varphi \phi^4 \chi$ terms}

These terms come from the variation of $\varphi$ in the potential
and from non-linear variation of $\psi$ in the last fermionic
item of (\ref{Act}):
\be
\sum_{K\neq I,J}%\left\{
2i(\bar\varepsilon \Gamma^K \chi)\varphi^K {\rm tr}
\left([\phi^I,\phi^J]\right)^2
+ i{\rm tr}\Big([\phi^I,\phi^J][\phi^L,\phi^M]\Big)
(\bar\varepsilon \Gamma^{KLM}\Gamma^{IJ}\chi)\varphi^K%\right\}
= \nn \\
= i\sum_{K\neq I,J}\varphi^K {\rm tr}\left\{[\phi^I,\phi^J]
\Big(2[\phi^I,\phi^J] (\bar\varepsilon \Gamma^K \chi) +
[\phi^L,\phi^M](\bar\varepsilon \Gamma^{IJK}\Gamma^{LM}\chi)
\Big)\right\}
\label{vara1}
\ee
How can these very different structures cancel each other?
The reason is that
antisymmetry in $KLM$ and in $IJ$ allows one to arbitrarily
change the order of $\Gamma$-matrices in $\Gamma^{KLM}$ and
$\Gamma^{IJ}$ and then Jacobi identity for the trace over
$\phi$-fields can be applied to ensure the cancelation.

Let $K=8$. Then there are three different
kinds of terms in remaining sum over $IJLM$.

$\bullet$ The pair $LM$ coincides with $IJ$, say, $I,J=1,2$ and
either $L,M=1,2$ or $L,M=2,1$.
These terms contribute:
$$
2[\phi^1,\phi^2]\Gamma^8 + [\phi^1,\phi^2]\Gamma^{812}\Gamma^{12}
+ [\phi^2,\phi^1]\Gamma^{812}\Gamma^{21} = 0$$
because $\Gamma^{812}\Gamma^{12} =
(-\Gamma^8\Gamma^2\Gamma^1)(\Gamma^1\Gamma^2) = -\Gamma^8$, while
$\Gamma^{812}\Gamma^{21} = +\Gamma^8$.

$\bullet$ Both indices $L$ and $M$ is different
from $I$ and $J$, say $I,J=1,2$ and $L,M=3,4$.
The corresponding contribution is obtained by reordering
commutators under the sign of trace:
$${\rm tr} \left\{\phi^4\left(
\Big[\phi^3,[\phi^1,\phi^2]\Big]\Gamma^{812}\Gamma^{43}
+ \Big[\phi^2,[\phi^1,\phi^3]\Big]\Gamma^{813}\Gamma^{42}
+ \Big[\phi^1,[\phi^2,\phi^3]\Big]\Gamma^{823}\Gamma^{41}
\right)\right\}  = 0
$$
due to Jacobi identity for the commutators, which can
be applied because all $\Gamma$-matrix structures are
the same: proportional to $\Gamma^{84}\Gamma^{123}$.
Note that one of the indices $L,M$ (but not $I,J$!)
could be equal to $K=8$.

$\bullet$ One of the indices $LM$ coincides with one of $IJ$,
another -- not, say $I,J=1,2$ and
$L,M=1,3$ or $I,J=1,3$ and $L,M=1,2$.
Such terms contribute
$$
{\rm tr}\Big([\phi^1,\phi^2][\phi^1,\phi^3]\Big)
\Gamma^{812}\Gamma^{13}
+ {\rm tr}\Big([\phi^1,\phi^3][\phi^1,\phi^2]\Big)
\Gamma^{813}\Gamma^{12}
= -{\rm tr}\Big([\phi^1,\phi^2][\phi^1,\phi^3]\Big)
\Gamma^{81}\Gamma^1\Big(\Gamma^3\Gamma^2 + \Gamma^2\Gamma^3\Big)
= 0
$$
because of anti-commutativity of $\Gamma$-matrices.
Once again, one of the indices $L,M$ could be $K=8$.

It is important here that the terms with $K=I$ or $J$
are excluded from the potential, because such terms
would contribute to the first item  in (\ref{vara1}), but
not to the second one (where $L$ and $M$ are different from $K$)
thus no cancelation occurs and supersymmetry would be broken.

\subsection{$\varphi^2 \phi^3 \psi$ terms}

The story about these terms is very similar:
they come from the variation of $\phi$ in the potential
and from non-linear variation of $\psi$ in the first bi-fermionic
item in the second line of (\ref{Act}):
\be
\sum_{K\neq I,J} 4i(\varphi^K)^2
{\rm tr} \Big([\phi^I,\phi^J][\phi^I,\bar\psi^A]\Big)
\Gamma^J_{A\dot A}\varepsilon^{\dot A} +
4i \varphi^I{\rm tr} \Big([\phi^J,\bar\psi^A]
[\phi^K,\phi^L]\Big)\varphi^M
\Gamma^{IJ}_{AB}\Gamma^{KLM}_{B\dot B}\varepsilon^{\dot B}
\label{vara2}
\ee
The second term can be rewritten as
$$
\varphi^I\varphi^M {\rm tr}\left(\bar\psi^A
\Big[\phi^J,[\phi^K,\phi^L]\Big]\right)
\Gamma^{IJ}_{AB}\Gamma^{KLM}_{B\dot B}
\varepsilon^{\dot B}
$$
Now we take into account Jacobi identity for commutators
and $I\leftrightarrow M$ symmetry.

$\bullet$ Let first $I\neq M$, say, $I,M=1,2$.
If all $\ J,K,L\neq 1,2\ $ then
$\ \Gamma^{1J}\Gamma^{KL2}+\Gamma^{2J}\Gamma^{KL1} = 0$.
If at least two of the three indices $J,K,L$ coincide
with $I$ or $M$, one of the two $\Gamma$-factors vanishes.
The only interesting case is when one of the three
coincides with $I$ or $M$, say, $J,K,L=1,3,4$
Then we get:

\centerline{ 
$
\varphi^1\varphi^2\left\{ \Big[\phi^1,[\phi^3,\phi^4]\Big]
\Big(\Gamma^{11}\Gamma^{342}+\Gamma^{21}\Gamma^{341}\Big) +
\Big[\phi^3,[\phi^1,\phi^4]\Big]
\Big(\Gamma^{13}\Gamma^{142}+\Gamma^{23}\Gamma^{141}\Big) +
\Big[\phi^4,[\phi^1,\phi^3]\Big]
\Big(\Gamma^{14}\Gamma^{132} + \Gamma^{24}\Gamma^{131}\Big)\right\}=
$} $$
= \varphi^1\varphi^2 \left\{
\Big[\phi^1,[\phi^3,\phi^4]\Big] -
\Big[\phi^3,[\phi^1,\phi^4]\Big] +
\Big[\phi^4,[\phi^1,\phi^3]\Big]\right\}\Gamma^{234} = 0
$$
because of Jacobi identity.

$\bullet$ Let now $I=M$. Then in the second term in (\ref{vara2})
all $J,K,L\neq I$, and it is equal to
\be
\sum_{I\neq J,K,L}(\varphi^I)^2{\rm tr}  \left(\bar\psi^A
\Big[\phi^J,[\phi^K,\phi^L]\Big]\right)
(\Gamma^{J}\Gamma^{KL})_{A\dot A}
\varepsilon^{\dot A}
\label{vara21}
\ee
If all the three indices $J,K,L$ are different, this sum
vanishes dues to Jacobi identity; if all the three coincide
than $\Gamma^{KL}$ is identical zero.
The only interesting cases
are $J=K$ and $J=L$, when
(\ref{vara21}) exactly cancels the first term in (\ref{vara2}).
Note that again it is important that terms with $K=I,J$
are excluded from the potential in (\ref{Act}).

\subsection{Terms with $DB$}

Such terms appear from variation of kinetic part of the
action in the first line of (\ref{Act}) and are compensated
by the variation of $A$-field in the Chern-Simons action:
both are proportional to the long derivative
$$\epsilon_{\lambda\mu\nu}(D^\mu B^\nu)_{i\bj} =
\epsilon_{\lambda\mu\nu}\Big(\partial^\mu B^\nu_{i\bj}
 + [A^\mu,B^\nu]_{i\bj}\Big)$$
Note once again that there are no $B^3$ terms in (\ref{Act}),
there is nothing to cancel their variation and they would
violate supersymmetry.
At the same time the $\varphi^2B^2$ term is present in the
first line of (\ref{Act}): as we discussed,
it plays the role in transition to $D2$ branes, but its
variation cancels against that of the $\psi B\chi$  and
does not produce $B^2$ terms. Only $DB$ is present in the
variation.

In a little more detail, the first line in (\ref{Act})
can be rewritten as
\be
-{\rm tr} (D_\mu\phi^I)^2 - 2\varphi^I {\rm tr}
(B^\mu D_\mu\phi^I) - (\varphi^I)^2{\rm tr}\, B_\mu^2 
+ i\,{\rm tr}\bar\psi^A \hat D\psi^A +
2i\,{\rm tr}(\bar\psi^A\hat B)\chi^A
\label{kite}
\ee
where $D_\mu\phi = \partial_\mu\phi + [A_\mu,\phi]$
is an ordinary adjoint long derivative with the gauge field 
$A_\mu$. The $B^2$ terms in the SUSY variation are
$$-2i\varphi^I (\bar\varepsilon\Gamma^I\chi){\rm tr}\, B_\mu^2
+2i\,{\rm tr} (\bar\varepsilon\Gamma^I\hat B)\hat B \chi = 0
$$
because  ${\rm tr}\, \hat B^2 = {\rm tr}\, B^2$.
Note that coefficient $2$ in front of the last terms in 
(\ref{kite}) -- and thus in the second long derivative in
(\ref{lode}), -- as well as anti-Hermiticity
(antisymmetry) of $B$ are important for this cancelation.

Similarly the variation of other terms in (\ref{kite})
provide terms with $F = \partial A+[A,A]$ and 
$DB=\partial B+[A,B]$ canceled respectively by the variations 
of $B$ and $A$ in the Chern-Simons term:
$\delta \int {\rm tr}FB = \int{\rm tr} 
(F\delta B - DB\delta A)$.

\subsection{Other terms}

The  terms which are cubic in fermion fields,
$\varphi\chi\psi^2$  and $\varphi\psi^3$, come from the
variation of gauge fields in the first line of (\ref{Act})
and from variation of scalars in bi-fermion terms in the
second line. Their cancelation requires adjustement of the
relative coefficient between the first and the second lines
of (\ref{Act}) and depends on the Fierz
identities for gamma-matrices.
The latter can be handled by making use of explicit 
$\Gamma$-matrix representation from s.\ref{fref}.
Most numerous are terms of the type $\varphi\phi^2\psi$,
they come from many places in (\ref{Act}) and we do not
analyze them in this paper.
Only when such analysis is completed one can take the
action (\ref{Act}) really serious.

\section{BL construction \label{BLa}}

In the remaining part of this paper we comment on the place
of (\ref{Act}) in the general BL theory.
As already mentioned in the introduction, (\ref{Act})
is associated with the special version (\ref{Nabra})
of the $3$-bracket -- the only one known for generic group
$U(N)$. It does {\it not} satisfy the antisymmetry requirement
for $f^{abcd}$ and original $SO(4)=SU(2)\times SU(2)$ example
of \cite{BL2} with $f^{abcd}=\epsilon^{abcd}$
is {\it not} the same as (\ref{Act}) for $N=2$.
The $SO(4)$ example is associated with the deformation of
(\ref{Nabra}) by addition of $\ I\cdot {\rm tr}(A[B,C])$,
with the unit matrix $I\in U(N)$,
which exists -- at least in such simple form -- only for $N=2$.

\subsection{Action and its invariance}

This section is a very brief summary of BL construction,
based on arbitrary (axiomatically defined) $3$-product.
The story begins at the end of s.\ref{fref}.
Note that we use the same notations $a,b,c,d$ and "tr"
as in ss.\ref{nalin},\ref{prepo}, but here they correspond
to a somewhat different group $\tilde G\neq G$:
the simplified BL action (\ref{Act}) with the gauge group $G=SU(N)$
is associated with $\tilde G = U(N)$.

\subsubsection{ Kinetic terms}

Let $\phi$ and $\psi$ belong to some {\it real} representation
$R$ of some gauge group $\tilde G$ with connection $\pA$.
Then short derivatives in (\ref{Lfree}) are substituted by
the long ones, and the $\tilde G$-invariant scalar product is involved
(denoted by "tr"):
\be
{\cal L}_{{\rm kin}} =
- {\rm tr}_R\ {\pD}^\mu\phi^I{\pD}_\mu\phi^I +
i{\rm tr}_R\ \bar\psi^A\hat{\pD} \psi^A
\label{Lkin}
\ee
Under the same transformation (\ref{phipsitra0})
we get:
\be
\delta S_{{\rm kin}} = \int \delta {\cal L}_{{\rm kin}} d^3x =
i\int {\rm tr}_R\
\bar\psi^A \pF_{\mu\nu}\sigma^{\mu\nu}\phi^I\Gamma^I_{A\dot A}
\varepsilon^{\dot A},
\label{varkin1}
\ee
(of course, $\varepsilon^{\dot A}$, is also an element of $R$).

\subsubsection{Chern-Simons self-interaction of the gauge field}

In order to compensate for this change one can add to
$S_{{\rm kin}}$ a Chern-Simons action for $\pA$,
\be
S_{{\rm CS}} = {\rm Tr} \int
\Big( \pA d \pA + \frac{2}{3} {\pA}^3 \Big)
= \epsilon^{\lambda\mu\nu} {\rm Tr}
\int \Big({\pA}_\lambda \partial_\mu {\pA}_\nu +
\frac{2}{3}{\pA}_\lambda {\pA}_\mu {\pA}_\nu \Big) d^3x
\ee
which varies as
\be
\delta S_{{\rm CS}} = \epsilon^{\lambda\mu\nu} {\rm Tr}
\int \pF_{\mu\nu}\delta {\pA}_\lambda
= \epsilon^{\lambda\mu\nu}
\int \pF_{\mu\nu}^{ab}\delta {\pA}_\lambda^{ab}
\ee
and can compensate the change in (\ref{varkin1}) if
(\ref{phipsitra0}) is complemented by
\be
\delta {\pA}_\lambda = -i\phi^I\otimes
\bar\psi^A \gamma_\lambda \Gamma^I_{A\dot A}\varepsilon^{\dot A}
\ \ \ \ {\rm or} \ \ \ \
\delta\pA^\lambda_{ab} = -i \bar\psi^A_a\gamma^\lambda
\Gamma^I_{A\dot A}\varepsilon^{\dot A} \phi^I_b
\label{Atra1}
\ee
Note that "Tr" is different from "tr": while the latter one
is for representation $R$ where $\phi$, $\psi$ and $\varepsilon$
belong, the former one is over representation $R\otimes R$,
where connection ${\pA}$ is taking its values.

\subsubsection{Twisted Chern-Simons}

The same result can be achieved for any action, which changes by
\be
\delta \tilde S =
\epsilon^{\lambda\mu\nu} {\rm Tr}
\int \pF_{\mu\nu}\delta {\npA}_\lambda
\label{varCS2}
\ee
provided
\be
\delta {\npA}_\lambda = -i\phi^I\otimes
\bar\psi^A \gamma_\lambda \Gamma^I_{A\dot A}\varepsilon^{\dot A}
\ \ \ \ {\rm or} \ \ \ \
\delta\npA^\lambda_{ab} = -i \bar\psi^A_a\gamma^\lambda
\Gamma^I_{A\dot A}\varepsilon^{\dot A} \phi^I_b
\label{Atra2}
\ee
An example is provided by "twisted Chern-Simons" action
\be
\tilde S_{{\rm CS}} = {\rm Tr} \int
\Big( {\npA} d {\pA} +
\frac{2}{3} {\npA}{\pA}^2 \Big)
= \epsilon^{\lambda\mu\nu} {\rm Tr}
\int \Big({\npA}_\lambda \partial_\mu {\pA}_\nu +
\frac{2}{3}{\npA}_\lambda {\pA}_\mu{\pA}_\nu \Big) d^3x
\ee
with
\be
{\pA}^{ab} = f^{abcd} {\npA}_{cd}
\label{pAnpA}
\ee

\subsubsection{Interaction terms}

Non-trivial transformation (\ref{Atra1})
of the ${\pA}$-field contributes new terms to
(\ref{varkin1}):
\be
{\rm Tr} \int \delta \pA_\mu \Big(
2\phi^I\otimes  {\cal D}^\mu \phi^I
- i\bar\psi^A\otimes\gamma^\mu\psi^A \Big) d^3x
\ee
These can be compensated by simultaneous addition of
interaction terms to the action  non-linear terms to the
SUSY transformation (\ref{phipsitra0}).
As shown in \cite{BL2} this can be done if (\ref{Atra2})
is used instead of (\ref{Atra1}), and all additions
are expressed in terms of the "structure constant"
$f^{abcd}$ from (\ref{pAnpA}), which can be used to define
a $3$-product
\be
R\otimes R\otimes R \rightarrow R:\ \ \
[X,Y,Z]^a = f^{abcd}X_bY_cZ_d
\ee
To raise and lower indices one also
needs a metric $h_{ab}$, which, however, does not show up
in the supersymmetry transformations.
In these terms
\be
{\cal L}_{{\rm int}} = \frac{1}{6}{\rm tr}_R
[\phi^I,\phi^J,\phi^K]^2 + \frac{1}{2}
{\rm tr}_R\ \bar\psi^A \Gamma^{IJ}_{AB}[\phi^I,\phi^J,\psi^B]
\ee
and
\be
\delta \psi_A = \hat\phi^I\Gamma^I_{A\dot A}\varepsilon^{\dot A}
+ \frac{1}{6}
[\phi^I,\phi^J,\phi^K]\Gamma^{IJK}_{A\dot A}\varepsilon^{\dot A}
\ee
The full action \cite{BL2}
\be
\int {\cal L}_{{\rm kin}}d^3x + \tilde S_{CS} +
\int {\cal L}_{{\rm int}}d^3x
\label{BLac}
\ee
is invariant, provided the $3$-product
satisfies the Jacobi-like "fundamental identity"
\cite{FI,G1,BL2,BL3}
\be
\Big[A,B,[C,D,E]\Big] =
\Big[[A,B,C],D,E\Big] + \Big[C,[A,B,D],E\Big]
+\Big[C,D,[A,B,E]\Big]
\label{fi}
\ee
and $f^{abcd}$ has certain symmetry properties.
It is usually required to be totally antisymmetric,
though this requirement can probably be relaxed \cite{G2}.

\subsubsection{Summary of transformations}

$$
\begin{array}{lc}
\delta\Big((\pD\phi^I)^2 + \bar\psi^A_a(\hat\pD\psi^A)_b
+ \tilde S_{{\rm CS}}\Big) &
\delta\pA_\mu^{ab}\left(\phi^I_a (\pD_\mu\phi^I)_b +
\bar\psi^A_a\gamma_\mu\psi^A_b\right)
+ \bar\psi^A_a\frac{\partial H^a_{A\dot A}}{\partial\phi^I_b}
(\hat\pD\phi^I)_b\varepsilon^{\dot A}
\\ & \\
\delta\Big( \bar\psi^A_a\psi^B_b
\Gamma^{IJ}_{AB}T^{IJ}_{ab}(\phi)\Big)& \!\!\!\!\!\!\!\!\!\!\!\!\!
+\bar\psi^A_a\psi^B_b\Gamma^{IJ}_{AB}
\frac{\partial T_{ab}^{IJ}}{\partial\phi^K_c}
\bar\psi^C_c\Gamma^K_{C\dot C}\varepsilon^{\dot C}
+ 2\bar\psi^A_a \Gamma^{IJ}_{AB}T^{IJ}_{ab}\Gamma^K_{B\dot B}
(\hat\pD\phi^K)_b\varepsilon^{\dot B}
+ 2\bar\psi^A_a\Gamma^{IJ}_{AB}T^{IJ}_{ab}H^b_{B\dot B}
\varepsilon^{\dot B}
\\ & \\
\delta \Big(V(\phi)\Big) &
+ \frac{\partial V}{\partial \phi^I_a}\bar\psi^A_a\Gamma^I_{A\dot A}
\varepsilon^{\dot A}
\end{array}
$$
There are three terms in the first line, three in the second and one
in the forth. Enumerating them, from 1 to 7, we have the following
cancelations to take place:
$$
\begin{array}{ccc}
2+4 &  \delta\pA_\mu^{ab}\bar\psi^A_a\gamma_\mu\psi^A_b +
\bar\psi^A_a\psi^B_b\Gamma^{IJ}_{AB}
\frac{\partial T_{ab}^{IJ}}{\partial\phi^K_c}
\bar\psi^C_c\Gamma^K_{C\dot C}\varepsilon^{\dot C} & = 0 \\ &&\\
6+7 &   2\bar\psi^A_a\Gamma^{IJ}_{AB}T^{IJ}_{ab}H^b_{B\dot B}
\varepsilon^{\dot B} +
\frac{\partial V}{\partial \phi^I_a}\bar\psi^A_a\Gamma^I_{A\dot A}
\varepsilon^{\dot A} & = 0
\\ && \\
1+3+5 \ \ \ & \delta\pA_\mu^{ab}\phi^I_a (\pD_\mu\phi^I)_b +
\bar\psi^A_a\frac{\partial H^a_{A\dot A}}{\partial\phi^I_b}
(\hat\pD\phi^I)_b\varepsilon^{\dot A} +
2\bar\psi^A_a \Gamma^{IJ}_{AB}T^{IJ}_{ab}\Gamma^K_{B\dot B}
(\hat\pD\phi^K)_b\varepsilon^{\dot B} & = 0
\end{array}
$$
Direct check of these cancelations involves application of
Fierz and other $\gamma$-matrix identities and is rather tedious.

\subsection{From $M2$ to $D2$ \label{MD2}}

A very important and instructive transition from BL
action for $M2$ branes in $d=11$ with $8$ transverse
fields $\phi^I$ to  $D2$ branes in $d=10$ with only $7$
transverse fields $\phi^i$ an the action
\be
\int d^3x \left( \sum_{i=1}^7 (D_\mu \phi^i)_a
(D^\mu \phi^i)_a
+ \frac{1}{4g_{YM}^2}\, {\rm tr}\, F^{\mu\nu}_aF_{\mu\nu}^a
+ {\rm fermions}\right)
\label{D2ac}
\ee
was considered in \cite{MP}.
Note that both $\phi^a$ and $A^a_\mu$ are now in the
same adjoint representation of the gauge group.

As explained in \cite{MP}, transition from $M2$ action
to $D2$ one in (\ref{D2ac}) takes place when one of the
scalar fields, associated with the $8$-th transverse
direction, acquires vacuum expectation value:
\be
<\phi^{I=8}_{a=0}> = g_{YM}
\ee
and $a$ in (\ref{D2ac}) takes one less value than $a$
in BL action (\ref{BLac}), so that
${\rm dim}(\tilde G) = {\rm dim}(G)+1$.
We label this extra value of $a$ in $\tilde G$ by $0$
(note that it is {\it not} the additional $0$-generator
of \cite{BL2}!).
Then for $a\neq 0$
$$
(\pD_\mu \phi^I)_a =
\partial_\mu\phi^I_a - \pA_\mu^{ab}\phi^I_b =
\partial_\mu\phi^I_a - f_{abcd}\npA_\mu^{cd}\phi^I_b =
\partial_\mu\phi^I_a - f_{abc0}\npA_\mu^{c0}\phi^I_b
- f_{a0cd}\npA_\mu^{cd}\phi^I_0
%= $$\vspace{-0.4cm}
%\be
=\partial_\mu\phi^I_a - f_{abc}A_\mu^{c}\phi^I_b
- B^a_\mu\phi^I_0
$$ %\ee
where
\be
f^{abc} = f^{abc0}
\label{ff}
\ee
are the ordinary structure constants of $G$.

The terms with $I\neq 8$ in (\ref{D2ac}) provide
\be
\sum_{i=1}^7
\left(\partial_\mu\phi^i_a + f_{abc}A_\mu^{c}\phi^i_b\right)^2
+ O(B\phi^2) =
\sum_{i=1}^7\Big((D_\mu\phi^i)_a\Big)^2 + O(B\phi^2)
\ee
to the action, while those with $I=8$ contribute
\be
g_{YM}^2B^a_\mu B^\mu_a
\ee
In combination with
\be
S_{CS} = \int (BF + B^3)
\ee
it provides the kinetic term
$\frac{1}{4g_{YM}^2}\int {\rm tr}\, F^2 d^3x$
for Yang-Mills field.

\bigskip

Note that the {\it number} of $B$-fields
\be
B_\mu^a = f_{a0cd}A_\mu^{cd}
\ee
is smaller than that of $A_\mu^{cd}$,
so that in the quantum case the measure
$[{\cal D}A]$ in functional integral would differ
from $[{\cal D}B]$ by non-trivial Jacobian factor.

\subsection{$3$-algebra and associated gauge fields
\label{3al}}

The key point of BL construction is the very interesting
$3$-algebra structure, supposedly generalizing the
Lie-algebra structure underlying the ordinary gauge symmetry.
BL construction associates with $3$-algebras a new kind
of gauge fields $A^{ab}_\mu$.
If the $\phi^I$-fields are interpreted as $X^I$,
describing $8$ transverse directions to the $M2$ brane
world volume in embedding $d=11$ space time, then
index $a$ is naturally treated as $a=(i\bj)$
where $i$ and $\bj$ label the copies of $M2$ brane and
associated string stretched between them.
Despite this interpretation is natural for $D2$ and not
obligatory to $M2$ branes, the reduction procedure
\cite{MP}, briefly described in the previous s.\ref{MD2},
strongly suggests that it also holds for $M2$.
Then $A^{ab}$ is actually carrying two pairs of indices
$A^{ab} = A^{i\bj,k\bar l} = A^{ij}_{kl}$
and looks like associated with a {\it pair of strings},
i.e. with a (fundamental) 2-brane stretching between the
two strings which stretch between the two pairs of $M2$ branes.
This is of course more than natural for the $M$-theory
and this is what makes the BL construction so attractive.
Unfortunately, {\it this} intriguing structure is fully
suppressed in the simplified action (\ref{Act}), where
gauge fields are the ordinary adjoints, associated with
pairs of branes rather than pairs of strings between them.

The big problem of BL construction is that no fully-non-trivial
examples of $3$-algebras are known.
Spectacular original $SO(4)$ example of \cite{BL2} appeared
to be difficult to generalize, despite certain effort
in \cite{G1}-\cite{MVR}.
Immediate idea about octonionic generalizations was shown
in \cite{BLS} to be not so straightforward.
Even the very inspiring reformulations of \cite{G1,BL2,G2}
do not provide immediate new examples.
One could hope that such examples are {\it obliged} to exist,
because $N$ branes with arbitrary $N$ certainly exist.
Unfortunately, eq.(\ref{Act}) demonstrates that these
configurations can probably be described by a very primitive
BL action, which is associated with degenerate $3$-bracket
(\ref{Nabra}), as we shall see in the next section.

Before proceeding to this last subject of the present paper,
it deserves emphasizing that the total-antisym\-metricity
requirement imposed in \cite{BL2}
over-constrains the possible $3$-algebra structures.
The $3$-bracket (\ref{Nabra})
corresponds to $f^{abcd}$ which is {\it not} totally antisymmetric,
in particular, $f^{0bcd}=0$ while $f^{abc0} = f^{abc}\neq 0$.
Only for $N=2$ it can be promoted to the totally antisymmetric
$f^{abcd} = \epsilon^{abcd}$ without violating the
"fundamental identity" (\ref{fi}).
Fortunately, as mentioned in \cite{G2},
the antisymmetry condition depends on the metric $h_{ab}$,
which does not appear in supersymmetry transformations
and does not affect the closure of the algebra.

%It also deserves mentioning that a
A way from Lie algebra structure
constants to generic $3$-algebras lies through solving
a {\it linear} equation for a {\it linear} operator
$t_{AB}$ \cite{G1,G2,MP}, $t_{AB}(P)=[A,B,P]$:
\be
t_{AB}([C,D])-t_{CD}([A,B]) =
\left[C,t_{AB}(D)\right] - \left[D,t_{AB}(C)\right]
\label{eqt1}
\ee
which should afterwards be constrained by a non-linear equation
\be
t_{AB}\Big(t_{CD}(E)\Big) - t_{CD}\Big(t_{AB}(E)\Big) =
t_{t_{AB}(C),D}(E) + t_{C,t_{AB}(D)}(E)
\label{eqt2}
\ee

\section{The simplified action (\ref{Act})
in the general BL framework \label{BLasi}}

The $U(N)$ Lie algebra decomposes as
\be
U(N) = U(1) + SO(N) + S(N)
\ee
(anti-Hermitian matrix is a sum of imaginary unit,
real antisymmetric and imaginary symmetric)
and this is a {\it symmetric} decomposition: $S(N)$ is
not a subalgebra of $U(N)$, only a representation of $SO(N)$, 
but with the special property
\be
[S(N),S(N)]\subset SO(N)
\ee
-- this is important for the algebraic discussion of \cite{G1}.
It is also important that in $U(N)$ one can multiply matrices,
not only commute (there are $d^{abc}$ constants in
addition to $f^{abc}$).
However, this possibility is actually not used in construction
of Nambu bracket (\ref{Nabra}) \cite{ALMY,G2}
\be
[A,B,C] = {\rm tr}(A) \cdot [B,C]
+ {\rm tr}(B)\cdot [C,A] + {\rm tr}(C)\cdot [A,B]
\label{Nabra1}
\ee
As already mentioned in the previous sections,
there is no term with $I$
(with non-vanishing trace) at the r.h.s. --
this makes the bracket a kind of degenerate and
$f^{abcd}$ very asymmetric.
In the case of $U(2)$ one can add $\ I\cdot {\rm tr}([A,B]C)$,
so that $[A,B,C]^d = \epsilon^{abcd}A_aB_bC_c$.
It is easy to see that the linear operator
\be
t_{AB}(*) = {\rm tr}(A)\cdot[B,*\,] - {\rm tr}(B)\cdot[A,*\,]
+[A,B]\cdot {\rm tr}(*\,)
\ee
which defines (\ref{Nabra1}), satisfies equations
(\ref{eqt1}) and (\ref{eqt2}).

It remains to rewrite the action (\ref{BLac}) in terms
of this $3$-product (taking care of the lack of total
antisymmetry).
We denote the extra $0$-components (traces) of the fields
$\phi^I_{i\bj}$ and $\psi^A_{i\bj}$ through
$\varphi^I$ and $\chi^A$, while the BL field ${\cal A}_{ab}
= {\cal A}_{i\bj,k\bar l} = {\cal A}^{jl}_{ik}$
is split into
${\cal A}^{jk}_{ik} = \frac{1}{2}A^j_i$ and
${\cal A}^{jk}_{ki} - {\cal A}^{kj}_{ik} = B^j_i$
(this is not a one-to-one change, as already mentioned in
the end of s.\ref{MD2}).
With these notations we get, for example, from (\ref{pAnpA}):
\be
\pA^{jp}_{iq}\phi^q_p =
\npA^{kj}_{kl}\phi^l_i - \npA^{kl}_{ki}\phi^j_l -
\npA^{jk}_{lk}\phi^l_i + \npA^{lk}_{ik}\phi^j_l +
\phi^l_l \Big(\npA^{jk}_{ki} - \npA^{kj}_{ik}\Big)
= [A,\phi]^j_i +  B^j_i \varphi
\ee
Expressions (\ref{lode}) for long derivatives are immediate
corollaries of this relation.
In the same way one can deduce from (\ref{BLac})
all other terms of the action (\ref{Act}).

\section{Conclusion}

To summarize, eq.(\ref{Act}) explicitly describes a
presumably-$OSp(8|4)$-invariant 
BL world-volume action for an arbitrary
number $N$ of coincident $M2$ branes.
Despite being based on a very primitive $3$-algebra,
associated with the $U(N)$ Nambu bracket (\ref{Nabra})
\cite{ALMY,G2}
and thus lacking the mysterious non-adjoint gauge fields
$A^\mu_{ab}$, this simplified action is sufficient to
describe transition {\it a la} \cite{MP}
to generic $D2$ branes with arbitrary gauge group.
Thus it can provide a far simpler and down-to-Earth
solution to the problem posed in \cite{JS} than the generic
BL construction, deeply hiding the underlying
$3$-algebra structure. As explained in s.\ref{3al} above,
this is rather a drawback than an advantage of eq.(\ref{Act}).

\section*{Acknowledgements}

This work  is partly supported by Russian Federal Nuclear
Energy Agency and Russian Academy of Sciences,
by the joint grant 06-01-92059-CE,  by NWO project 047.011.2004.026,
by INTAS grant 05-1000008-7865, by ANR-05-BLAN-0029-01 project,
by RFBR grant 07-02-00645 and
by the Russian President's Grant of Support for the Scientific
Schools NSh-3035.2008.2

\end{document}